# Laser polarization as a critical factor on the SERS-based molecular sensing performance of nano-gapped Au nanowires

Simón Roa[1,2]*, Terunori Kaihara[3], María Laura Pedano[1,2,4]*, Henrik Parsamyan[5], Paolo Vavassori [3,6]

[1] Instituto de Nanociencia y Nanotecnología (CNEA - CONICET), Nodo Bariloche, Av. Bustillo 9500, (8400) S.C. de Bariloche, Río Negro, Argentina.
[2] Laboratorio de Fotónica y Optoelectrónica, Centro Atómico Bariloche, Av. Bustillo 9500, (8400) S. C. de Bariloche, Río Negro, Argentina.
[3] CIC nanoGUNE BRTA, Tolosa Hiribidea, 76, 20018 Donostia-San Sebastián, Spain.
[4] Instituto Balseiro, CNEA-Universidad Nacional de Cuyo (UNCUYO), Av. E. Bustillo 9500, R8402AGP S. C. de Bariloche, Río Negro, Argentina.
[5] Institute of Physics, Yerevan State University, 1 Alex Manoogian, Yerevan 0025, Armenia.
[6] IKERBASQUE, Basque Foundation for Science, Plaza Euskadi 5, 48009 Bilbao, Spain.

*Corresponding authors e-mails: sroadaz8@gmail.com (S. Roa), ml.pedano@cab.cnea.gov.ar  (M. L. Pedano).

**Abstract**

Nowadays, Au dimer-based nanostructures are exhaustively studied due to their outstanding potential as plasmonic nanoantennas for future applications in high-sensitivity molecular sensing by Surface-Enhanced Raman Spectroscopy (SERS). In this work, we analyze nano-gapped Au nanowires (NWs) or Au-NWs dimers for designing efficient nanoantennas, reporting an exhaustive study about dimer length and laser polarization orientation effects on their SERS-based molecular sensing performance. Arrays of nanoantennas with gaps about 22 ± 4 [nm], nominal square cross-sections of 60 [nm] x 60 [nm], and different lengths from 300 [nm] up to 1200 [nm] were fabricated by Au evaporation and subsequent e-beam lithography. SERS performance was studied by confocal Raman microscopy using a linearly-polarized 633 [nm] laser. A critical impact of the polarization alignment on the spectral resolution of the studied Raman marker imprint was observed. Results show that the Raman signal is maximized by aligning the polarization orientation with the nanowire long axis, it is reduced by increasing the relative angle, and it is abruptly minimized when both are perpendicular. These observations were consistent with numerical simulations carried out by the FDTD method, which predict a similar dependence between the orientation of linearly-polarized light and electric-near field amplification in the nano-gap zone. Our results provide an interesting paradigm and relevant insights in determining the role of laser polarization on the Raman signal enhancement in nano-gapped Au nanowires, showing the key role of this measurement condition on the SERS-based molecular sensing efficiency of this kind of nanostructure.

**Keywords:** Surface-Enhanced Raman Spectroscopy, Molecular Sensing, Plasmonic Nanoantennas, Au Nanowires, Electric Near-Field Enhancement.

## 1. Introduction

Since the last two decades, Plasmon-Enhanced Raman Spectroscopy (PERS), which includes Surface-Enhanced Raman Spectroscopy (SERS), Shell-Isolated Nanoparticle-Enhanced Raman Spectroscopy (SHINERS), and Tip-Enhanced Raman Spectroscopy (TERS), has been established as a powerful technique to characterize the Raman spectral footprint of different kinds of surface species with high chemical sensitivity and spatial resolution [1]. PERS can provide information on the chemical identity of specific analytes with sensitivities down to the single-molecule level and with sufficient spatial resolution to observe individual vibrational modes. Thus, PERS has found applications in diverse areas that require trace detection of pollutants, viruses, bacteria, cancerous cells, polymers, heavy metals, and many other types of inorganic and organic materials [1 – 7]. This detection versatility has enabled applications from academic to industrial level, as well as in satisfying daily society demands like disease diagnostic in medicine and food/water quality control in agriculture [8 – 14].

In particular, SERS has become widely applied for high-sensitivity molecular and biological sensing due to its versatility in using a wide variety of nanostructured materials with outstanding optical properties, which can vary from roughened thin films to more complex nanostructures (porous surfaces, nanoparticles assemblies, nanocavities, nanowires-based dimers, photonic crystals, etc.) [8, 15 – 20]. Initially, plasmonic metal nanoparticles have been employed as optical nanoantennas due to the presence of Localized Surface Plasmon Resonance (LSPR) modes that arise from the non-propagating collective oscillations of free charge carriers in metal/dielectric interfaces [21, 22]. Nanoparticles with LSPR states possess large optical cross-sections and strong electric field enhancement, enabling them to be applied for SERS-based chemical and biological sensing [8]. These nanostructures can transfer far-field radiation to near-field energy (and vice versa), making them very efficient for nanoscale-confinement of light and induction of strong electric-near field *hot spots* (zones with high electric field amplification). Thus, nanoparticles have been introduced as a bridge to connect the gap between the macroscopic and the microscopic scale for light manipulation [23].

In addition to LSPR modes of metal nanoparticles, Surface Plasmon Polaritons (SPP) supported in metallic nanowires or nanoslits also enable light confinement in nanoscale regions [24, 25], where the near-field interaction between these kinds of nanoantennas and fluorescent emitters or Raman molecules generates highly enhanced directional light emission and enhanced Raman signals [26]. Therefore, these plasmonic nanoantennas have been widely studied for achieving efficient control of the light direction and electric near-field amplification at the nanoscale [26]. In addition to applications in SERS-based molecular sensing [26 – 29], these capacities have led plasmonic nanoantennas to be applied in other fields such as photocatalysis [30], solar energy harvesting [31], nonlinear optics [32] and active color switching [33]. To design nanoantennas, many ideas have been borrowed from conventional macroscopic schemes. One of the most famous examples is Yagi-Uda antennas, which have been widely used in analog television, shortwave communication links, radar antennas, and broadcasting stations [34]. As the nanoscale counterpart, Yagi-Uda nanoantennas can redirect light emission in a desired direction with a narrow angle and enhanced signal intensity in a nanoscale region [35, 36].

Nano-Gapped Metallic Nanowires (NGMN) have been proposed as efficient Yagi-Uda-like nanoantennas, presenting unique plasmonic properties to support propagating SPP states that can induce LSPR states confined at the gap region under determined size and shape conditions, and producing electric field *hot spots* with exceptionally high Electric-Near Field Enhancement (ENFE) factors **[37, 38]**. In SERS context, this phenomenon enables a strong light-matter (analyte) interaction at the gap region and consequently enhances Raman signals from the molecule **[38 – 42]**. The nanogap junction could afford extraordinary ENFE, even high enough for single-molecule detection **[41 – 43]**. Historically, Au and Ag have been preferred for the design of nanoantennas due to their low chemical reactivity and capacity to induce intense electromagnetic field amplifications by excitation of LSPR modes and enhanced Raman signals when excited by visible or near-infrared light, enabling non-destructive molecular sensing for a wide variety of inorganic and organic compounds **[44, 45]**.

SERS performance of Au-based NGMN was first studied experimentally by M. L. Pedano *et al* **[37]** and then by other authors **[38, 40 – 42]**. Yet, the effects of light polarization on this aspect have not been systematically explored experimentally. It is known that light polarization can be a determining factor on the ENFE, especially in nanoantennas with high degrees of shape anisotropy like nanowires that lead to the emergence of preferential SPPs states along different anisotropy axes **[27]**. Non-linearly polarized light is usually used to excite as many plasmonic modes as possible to ensure good SERS signals **[1]**. However, this practice is efficient when using nanostructures like roughened metallic surfaces or randomly distributed nanoparticles, where the shape, orientation, and spatial distribution of the scattering centers are arbitrarily spread through. In this kind of scenario, the excitation of surface plasmons with non-linearly polarized light makes sense since it is more efficient for SERS to have contributions of plasmon resonances induced in as many scattering centers as possible.

For the case of nanostructures with a high degree of shape anisotropy like NGMN, the application of non-linearly or randomly polarized light is not the most efficient option due to its weak electric field coupling with the preferential plasmonic modes propagating along the main axis of the nanostructure given by the shape anisotropy **[27]**. This kind of nanowire-based antenna is mainly characterized for presenting a strong propagating dipole resonance mode along the wire long axis, which is responsible for high electric-near field enhancements at the nanogap region. The criteria for incident light to efficiently excite this resonance mode are to linearly polarize light with the electric field vector parallel to the length of the nanowire, and also vertical illumination. Aligning the light polarization vector with the nanowire length (long axis) ensures a strong coupling between the light electric field and free electrons of the metal that support propagating oscillations along this length (fundamental SPP state or dipolar mode), which induce LSPR states confined at the gap region and produce electric field "hot spots" with exceptionally high ENFE **[27]**. The significant advantages of these incidence conditions include very high optical absorbance, high electric-near field intensity, strong interlayer coupling effect, and larger area "hotspot" **[41]**.

An early experimental approach to polarization effects in NGMN was given by H. Wei *et al* **[46]** who studied the polarization dependence of SERS in coupled gold nanoparticle-nanowire systems. A similar study, but considering AlN nanowires, was also carried out by Hsu *et al* **[47]**. In both cases, it was observed that the Raman signal enhancement was maximized when the polarization was aligned with the nanowire long axis. In our case, we report a systematic and exhaustive study, both theoretical and experimental, about laser polarization effects on the SERS sensing efficiency, or Raman signal enhancement, of nanoantennas based on nano-gapped Au nanowires. Our results show that the Raman signal enhancement is strongly dependent on the polarization alignment concerning the nanowire long axis, being consistent with Finite-Difference Time-Domain (FDTD) simulations that predict a similar behavior for our particular system. This research shows the critical impact of the measurement configuration on the SERS-based molecular sensing efficiency, determined by the role of the laser polarization alignment on the Raman signal enhancement in the nano-gapped Au nanowires.

## 2. Experimental details

**2.1. Fabrication of nanoantennas based on nano-gapped Au nanowires.** Nano-gapped Au nanowires were fabricated by Au evaporation, negative e-beam lithography, and Reactive Ion Etching (RIE), as shown in the scheme of **Fig. 1(a)**. The samples were fabricated on a 7 [mm] x 7 [mm] Si substrate following the procedure outlined in this figure. After cleaning the Si substrate with acetone, isopropanol and $O_2$ plasma, a 3 [nm]-thick Ti adhesion layer was deposited by e-beam evaporation (evaporation rate: 0.5 Å/s), followed by a 60 [nm]-thick Au film (evaporation rate: 0.8 [Å/s]). A negative tone resist (ma-N 2401) was then spin-coated onto the substrate at 4000 rpm for 60 seconds (~ 100 [nm] of nominal thickness).

The nanowires patterns were defined on the resist using an e-beam lithography system (RAITH150 Two) with an acceleration voltage of 20 [kV]. Due to the intrinsic characteristics and fabrication conditions of this method, the overall geometry of the obtained nanoantennas is similar to rectangular. They consist of two facing rectangular Au nanowires with length ($L$) for each segment, and a nominal square cross-section (area = $d$ x $d$) of 60 [nm] x 60 [nm], where $d$ is the nanowire thickness, separated by a nanogap ($g$) of 22 ± 4 [nm]. This gap size was chosen following previous similar research **[37, 38]** that selected it to fit an oligonucleotide of that length to bridge the gap for future biosensing analysis and molecular conductivity studies towards molecular electronic applications. An array of 6 x 6 sets of nanowires with different $L$ nominal values was fabricated on the Si substrate, as shown in **Fig 1(b)**. Each set contains a total of 5 x 5 nanowires with same nominal $L$ value. The $L$ values were systematically varied from 300 [nm] to 1200 [nm] in increments of 25 [nm] for each different set. Special care was taken to optimize the pattern biasing and electron dose scaling to achieve the desired nanowire geometries. After the e-beam exposition, the exceeding resist was developed by substrate immersion in an aqueous alkaline-based developer (ma-D525). Then, the pattern was transferred to the metal layers through Ar⁺ ion milling (4Wave IBE20L01) with a beam voltage of 300 [V], a beam current of 50 mA, and an acceleration voltage of 50 [V]. Finally, the residual resist was removed by immersion in N-Methyl-2-pyrrolidone (NMP) solution, leaving the desired nano-gapped Au nanowires.

**2.2. Confocal Raman spectroscopy.** Methylene Blue (MB) was employed as a Raman active analyte for confocal Raman spectroscopy studies of the SERS sensing efficiency of our samples. MB is generally used due to its simple implementation for surface functionalization and easily identifiable Raman spectrum **[48]**, making it a common Raman marker to test the electric field enhancement properties of other nanostructures and ideal for comparison purposes. For these reasons, MB was selected to assess the performance of our nanoantennas for SERS-based molecular sensing. In addition, as analyte of interest by itself, MB is commonly used in medical practice as a dye in microbiological staining and drug in treatments for Methemoglobinemia, dermatological diseases, etc **[49 – 51]**. It is also a common water pollutant derived from textile and food industry processes, and its detection is a key issue in water quality controls **[52 – 54]**. Due to its biological and industrial relevance, much attention has been focused on developing detection methods for MB.

For the NWs surface functionalization, 2 [μL] of a 10 [μM]-MB ethanolic solution was used to cover the entire 7 [mm] x 7 [mm] substrate containing the nanoantennas arrays and then dried at ambient conditions (without rinsing) to finalize the functionalization with MB molecules. Raman spectroscopy studies were carried out immediately after this process. Assuming a homogeneous distribution of the MB solution (3.2 [ng/μL], considering an MB molar mass of 319.85 [g/mol] [55]), an MB surface mass density of approx. 0.13 [ng/mm²] or 13 [ng/cm²] was estimated for each sample surface. LabRAM HR Evolution Confocal Raman Microscope (from HORIBA, Ltd.) was used to perform micro-Raman spectroscopy measurements.

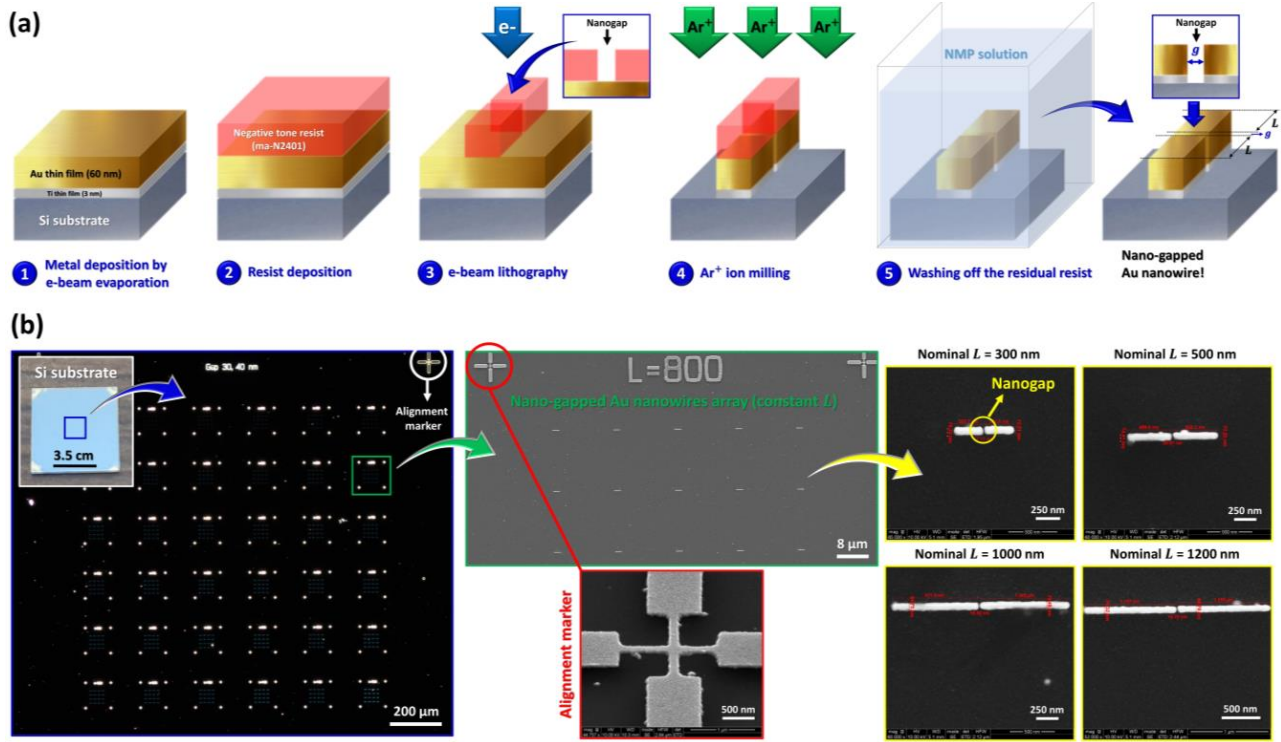

**Fig. 1. (a)** Idealized scheme of the fabrication steps carried out to fabricate the nano-gapped Au nanowires by e-beam lithography. **(b)** Different magnifications of the fabricated nanowires array on a 7 [mm] x 7 [mm] Si substrate. The studied sample consists of an array of 6 x 6 sections with different $L$ values from 300 to 1200 [nm] with a step of 25 [nm] between each section. Each $L$-constant section contains an array of 5 x 5 nano-gapped Au nanowires.

Raman spectra were measured in different samples by the punctual acquisition mode at the nanoantenna gap zone, using a 633 [nm] linearly-polarized laser with a 100X microscope objective (spot size of approximately 2 [μm] in diameter). Measurements were performed using a laser power of 0.4 [mW] (10 % of maximum output power). A Raman shift range of 1050 – 2100 [cm$^{-1}$], a 600 [gr/mm] grid, 10 accumulations and 3 [s] of acquisition time per accumulation were set for each spectrum measurement. When performing punctual Raman measurements, the best approach to positioning the laser spot on the nanogap zone is using the reference coordinate system given by the optical microscope as shown in **Fig. 2**. The origin of the coordinate system is positioned as close as possible to the nanogap zone, which can be easily detected by the optical microscope, and then punctual measurements are performed. During the measurements, a sub-micron position drift can occur. So, a representative number of measurements were done to overcome this effect on the average Raman response. In our case, Raman spectra were taken in 15 different samples of the same nominal $L$ at the nanoantenna gap zone, observing just slight fluctuations in Raman response as shown in the figures of **Appendix A**. This demonstrates that uncertainties in the laser positioning on the nanogap is a secondary-order parameter, not affecting considerably the Raman response deviations, and showing that the enhancement of Raman scattering signal at the nanogap is predominant over any small deviation of the laser spot with respect to the gap location.

Sub-micron Raman mapping studies (spatial resolution of 100 [nm])) were carried out on Au NWs to sense the complete SERS emission intensity and distribution with respect to the nanoantenna geometry, and compare it with the general behavior predicted by the simulations. DuoScan™ Imaging module of the LabRAM spectrometer was used for this task to minimize external vibrations in the sample holder stage and to ensure good stability of the laser spot position during the spectra acquisition and Raman mapping at the sub-micron scale. This technology is based on a combination of scanning mirrors that raster the laser beam across a pattern chosen by the operator with reliable sub-micron positioning and a maximum spatial resolution of 50 [nm] (DuoScan™ - HORIBA), which is critical when analyzing nanoantennas. In practice, this resolution is guaranteed. In our case, we limited to 100 nm because of the considerably higher acquisition times required to compensate the Raman intensity decay due to pixel size reduction (50 nm). Large-scale mapping areas (6 [μm] x 4 [μm]) concerning nano-gapped Au NWs dimensions were chosen to avoid potential random position shifts (typically from hundreds of nanometers to a few microns) of the initially defined scanning area. This minimizes the positioning errors of Raman mapping measurements, which could lead to get Au NWs off the scanned area. Once the image is obtained, the correspondence of the Raman map with the Au NW optical image is analyzed.

The effects of rotating the laser polarization in an angle $\theta$ concerning the nanowire long axis were studied by using an experimental configuration similar to that proposed by HORIBA, Ltd (HORIBA Scientific - Raman Spectroscopy), (see scheme of **Fig. 2(b)**). In the current case, a laser polarization rotator ($\lambda/2$ retarder) was put at the laser path before entering the spectrometer optics and a Raman polarization analyzer was put at the laser path just before the spectrograph grating and CCD. Thus, different Raman spectra were measured varying the angle $\theta$ from 0 to 360 [deg] and using the same acquisition conditions previously mentioned in each case.

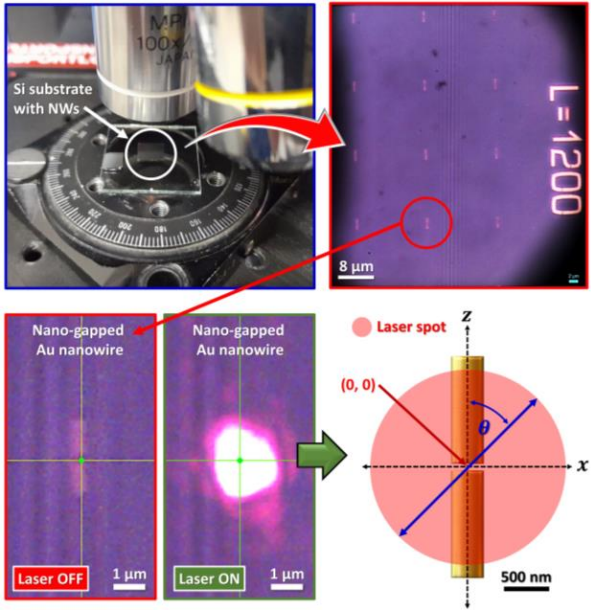

**Fig. 2.** Optical images taken with the LabRAM microscope of a nanowires array. The ideal scheme at the bottom right represents the experimental configuration used to measure the effects of rotating the laser polarization (represented by the blue line) concerning the nanowire long axis (aligned with x-axis) in an angle $\theta$. The blue line represents the linear polarization of the laser.

## 3. Results and discussion

**3.1. Numerical studies on ENFE in nano-gapped Au nanowires.** The interaction between light and surface plasmons can generate SPPs modes propagating along the nanowire that result in allowed standing waves, which can lead to the emergence of strong LSPR states at the nanogap that produce ENFE "hot spots" when tuning the adequate $L$ conditions. Results obtained in previous works for this kind of nanostructures [37, 38, 56, 57], but considering cylindrical wires, have demonstrated the periodic dependence between the SERS intensity at the nanogap and the $L$ parameter in agreement with ENFE calculations. This knowledge and our own experience have demonstrated that the optimum ENFE strongly depends on the precise control of geometric parameters. So, if the synthesis method is changed (in this case e-beam lithography instead of on-wire lithography [37, 38]) and the geometry of the nanoantenna is varied (now rectangular rather than cylindrical [37, 38, 56, 57]), new simulations must be carried out to assess the adequate parameters to control during the experimental process and predict which ENFE can be expected according to the limitations of the fabrication procedure. Therefore, to have an idea about the adequate design dimensions of our samples for future fabrication, we resorted to numerical methods to simulate and study the effects of different size parameters ($L$, $d$ and $g$) on the ENFE at the nano-gap region in rectangular nanoantennas.

In our case, COMSOL Multiphysics software – Wave Optics module was used to simulate the electric-near field spatial distribution. A 633 [nm] linearly-polarized Gaussian beam (radius = 1 [μm]) was used as excitation source in main simulations. This excitation wavelength was selected considering the value that ensures the best electric-near field enhancement factor in the visible range for the proposed nanostructures, which is associated with the main dipolar plasmonic resonance. This resonance condition was suggested by preliminary simulations (see **Fig. B** in **Appendix B**), which predict the highest enhancement factors in our nano-gapped Au nanowires for wavelengths between 625 and 640 [nm]. The beam incidence was considered to be perpendicular to the nanowire long axis and its symmetry axis perfectly aligned with the gap center, as shown in **Fig. 3(a)**. As stated in section 2.1, nanowires were considered with a square cross-section (area = $d \times d$), length $L$ per segment and gap $g$. In all cases, $g$ = 24 [nm] was set following previous similar research [56, 57], and considering the typical exponential growth of SERS signal amplification at the gap zone from $g < 40$ [nm] in these kinds of nanostructures [58]. $L$ and $d$ were varied in ranges of 20 – 2000 [nm] and 60 – 160 [nm], respectively.

To characterize the efficiency for SERS applications of these kind of nanoantennas, the Enhancement Factor ($EF$) figure of merit ($g^4 = |E|^4/|E_0|^4$) [59, 60] was used, where $E$ is the local electric field and $E_0$ is the amplitude of the incident electric field. In our case, we determined the effective $EF$ ($EF'$) of a system with particular dimensions as an average of the $EF$ values calculated over a square surface $d^2$ (perpendicular to the nanowire long axis), centered at the gap region (see the inset figure in **Fig. 3(b)**) as follows:

$$EF = \frac{\int |E|^4/|E_0|^4 \, dS}{\int dS}, \qquad (1)$$

where $dS$ is the differential surface area. The analyzed area $d^2$ was discretely divided in 1 [nm$^2$]-area square pixels ($dS$) for simulations. Thus, the $EF'$ can be simply estimated as:

$$EF' = \frac{1}{d^2}\sum_{i=1}^{d^2} |E_i|^4/|E_0|^4, \qquad (2)$$

where the sub-index $i$ represents the $i$-th pixel. **Fig. 3(a)** shows the evolution of the $EF'$ magnitude concerning different $L$ values for nano-gapped Au nanowires with $d$ = 60, 80, 100, 120, 140 and 160 [nm]. In some graphs, inset color maps show the distribution of the normalized electric field magnitude ($E/E_0$) at the nanogap center in the $x - y$ plane (parallel to the nanowire lateral cross-section) at different resonant modes for determined lengths $L$. As originally observed by M. L. Pedano *et al* [37] and reported in similar systems by her collaborators [57, 58], a strong ENFE can be achieved at the gap zone by choosing the adequate $L$ conditions. Simulations presented in **Fig.3 (a)** indicate that, compared to $L_0$, the $L_1$ value maintains a similar $EF'$ intensity for $d$ = 80 – 100 [nm], or slightly smaller for $d$ = 60 [nm] but still higher than those observed for larger $d$ magnitudes. On the other hand, for $d \geq 120$ [nm], $EF'$ values associated with $L_1$ are always significantly smaller than those observed for $L_0$. The first $EF'$ maximum peak is associated with the dipole resonance while the others are expected for multipole resonances [38].

Simulations suggest that the most efficient size condition that maximizes the enhancement factor corresponds to nanowires with $d$ = 60 [nm], which has been selected as a future fabrication parameter. On the other hand, deviations from the ideal proposed nanowire geometry, for example, curved edges at the gap zone, can considerably influence the ENFE magnitude. This fact is illustrated by simulations carried out in systems with $d$ = 120 [nm] for different curvature radii ($r$ = 5 and 10 [nm]) of the nanowire edge at the gap zone, as shown in **Fig. 3(b)**. In this case, we can observe a considerable decay of the ENFE at the gap zone just considering a reasonable curvature radius of $r$ = 5 [nm]. For higher curvatures, the ENFE decays even more, blurring the dependence of this factor with $L$ and the emergence of strong resonance modes as observed in cases with sharp geometry. Thus, the experimental Raman enhancement of the fabricated samples could not necessarily show such a strong and clear $L$-dependence on the enhancement peaks, as predicted by idealized simulations.

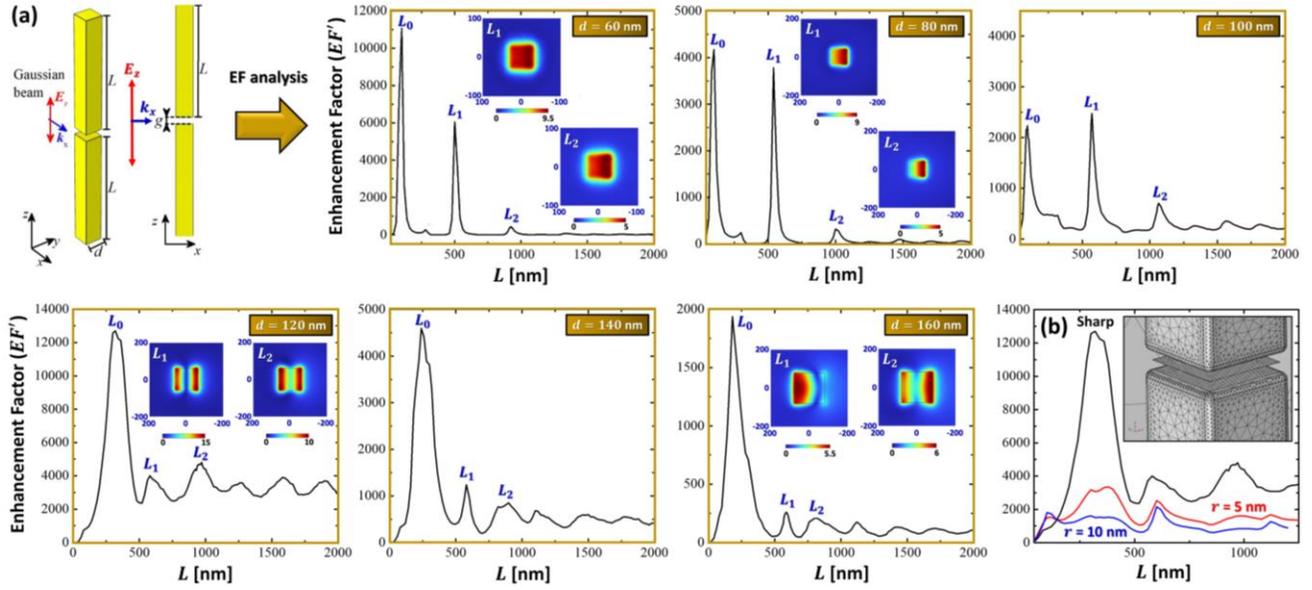

**Fig. 3. (a)** Evolution of the $EF'$ magnitude concerning different $L$ values for nano-gapped Au nanowires with nanowire thickness $d$ = 60, 80, 100, 120, 140 and 160 [nm] for an excitation wavelength of 633 [nm]. The inset color maps show the distribution of the normalized electric field magnitude ($E/E_0$) at the nanogap center in the $x-y$ plane (parallel to the nanowire lateral cross-section) at different resonant modes for determined lengths $L$. **(b)** Effects of different curvature radius ($r$ = 5 and 10 [nm]) of the nanowire edge at the gap zone for a system with $d$ = 120 [nm] and $L$ = 20 – 1250 [nm]. The inset figure helps to visualize the magnitude of the considered curvature concerning the nanowire dimensions and the location of the integration plane typically considered for calculating $EF'$.

**3.2. Experimental impact of the $L$ parameter on the Raman enhancement of nano-gapped Au nanowires. Fig. 4(a)** shows different Raman spectra of MB measured for $\theta = 0$ [deg] in nano-gapped Au nanowires with lengths $L$ from 300 up to 1200 [nm], observing the impact of this parameter. Considering possible fabrication deviations from nominal features and to have a reliable statistical representation, each spectrum shown for a particular $L$ value corresponds to an average of 15 measurements carried out in 15 different randomly chosen nanowires. Each measurement was carried out with a total acquisition time of 30 seconds, according to the experimental conditions stated in section 2.2. In general, we observed slight dispersion in the Raman intensity of spectra measured in samples with equal nominal $L$ values, as shown in **Fig. A** of **Appendix A**. Nevertheless, this experimental dispersion is small enough to observe differences in the SERS performance of samples with different nominal $L$ values. In a quick view of **Fig. 4(a)**, the Raman signal magnitude seems to be slightly dependent on $L$ variations. All the peaks observed in each Raman spectrum are associated with the MB chemical footprint, being in good agreement with other spectra reported for this compound **[53, 61 – 63]**. In particular, we can observe the most intense peak at 1621 [cm$^{-1}$] which is associated with the main vibration band of MB related to the C-C ring stretching **[61, 62]**. Other main peaks like those observed at 1300 [cm$^{-1}$] and 1396 [cm$^{-1}$] are associated, respectively, with the characteristic in-plane ring deformation of C-H bonds and the symmetrical stretching of C-N bonds in the MB molecular structure **[53, 63]**. These peaks and further ones associated with the typical MB vibrational spectrum can be better observed in the magnified spectra shown in **Fig. 4 (b)**. **Fig. 4 (c)** illustrates a submicron Raman map for a nanoantenna with nominal $L$ = 1000 [nm] to exemplify the observed ENFE amplification at the "hot spot" formed only at the nanogap.

To understand the quantitative effects of the $L$ parameter on the effective SERS efficiency of our nano-gapped Au nanowires, it was established the following criterion to quantify the relative or normalized SERS Enhancement Factor ($EF_{SERS}$) from the experimental Raman Intensity ($I_L$) measured at determined Raman shift ($\Delta\tilde{\nu}$) for each $L$:

$$EF_{SERS}(L) = \frac{I_L(\Delta\tilde{\nu}_{max}) - I_L(\Delta\tilde{\nu}_{ref})}{I_{L=500\ nm}(\Delta\tilde{\nu}_{max}) - I_{L=500\ nm}(\Delta\tilde{\nu}_{ref})}, \qquad (3)$$

where $I_L(\Delta\tilde{\nu}_{max})$ corresponds to the Raman intensity measured for the most intense peak (at $\Delta\tilde{\nu}_{max}$ = 1621 [cm$^{-1}$]) for any $L$ value. $I_L(\Delta\tilde{\nu}_{ref})$ corresponds to the Raman intensity measured for an arbitrarily-chosen reference position, $\Delta\tilde{\nu}_{ref}$ = 1590 [cm$^{-1}$], which is used as "baseline" intensity. **Eq. (3)** is useful to quantify the SERS enhancement induced by $L$ effects relative to the lowest one observed for the normalization value at $L$ = 500 [nm], such that $EF_{SERS}(L = 500\ [nm]) = 1$ and $EF_{SERS}(L \neq 500\ [nm]) > 1$.

**Fig. 4(d)** shows the calculated $EF_{SERS}(L)$ for $L$ between 300 and 1200 [nm], observing a nontrivial evolution of this factor concerning the $L$ variations. In particular, we can observe the presence of enhancement peaks with a well-defined periodicity about of 150 [nm], which apparently could not be explained from our numerical simulations. However, while the nominal cross-section of our nanowires was set to be 60 [nm] x 60 [nm], SEM images have shown that they are closer to 74 $\pm$ 3 [nm]. If we look for the expected enhancement peaks from **Fig. 3(a)** for $d$ = 80 [nm], we realize that there is a secondary peak about of $L$ = 1020 [nm]. In this sense, this could explain the highest $EF_{SERS}$ value observed experimentally for our nanostructures at $L$ = 1000 [nm]. The numerically-predicted $L$-dependent $EF$ periodic behavior could be blurred from theoretical models by the presence of geometrical deviations, such as significant edges "roundness" (high curvature radius) at the nanogap interface as predicted by the simulations shown in **Fig. 3 (b)**, or surface roughness not considered in the current simulations. SEM images presented in **Fig. 1(b)** for different nanowires clearly show the presence of these geometrical deviations. Thus, the behavior observed in our case could be strongly influenced by these kinds of defects. Additionally, the fact of observing good Raman signal enhancement for any $L$ value (as shown in **Fig. 4(a)**), despite numerical results that suggest zero enhancement in determined $L$ conditions for $d \leq 80$ [nm] (see **Fig. 3(a)**), could be indicating the presence of second order enhancement mechanisms. Metallic thin films fabricated by e-beam evaporation typically present nanogranular structures with nanometric roughness, which could produce a roughness-induced Raman scattering enhancement effect. AFM studies of surface topography and roughness have demonstrated the presence of nanometric roughness, as shown in **Appendix C**. Nanometric roughness may result in high electric-near field enhancements due to nanocavities or sharp edges-induced amplification [64 – 68]. Moreover, gap roughness can outcome in smaller ENFE and small changes in the optimum $L$ for the highest intensities, compared to smooth gap surfaces [38]. These effects were explained by S. Li *et* al [38]. For smooth-gapped nanoantennas, if the light is polarized along the longitudinal axis of the segments, only odd-order ($n$ = odd) resonance modes can be excited [69, 70] and the $n > 1$ resonance occurs at a smaller segment size, reflecting differences in coupling strength across the gap. However, gap roughness induces the reflection dephasing of the SPPs from the gap, which weakens and broadens the resonances, leading to more localized hot spots at the gap but weaker coupling across the gap [71]. Small peaks additional to the main resonance appear in rough-gapped nanoantennas attributed to even mode resonances. These modes cannot be excited in isolated smooth rods with light polarization along longitudinal axes, but they appear in rough gaps because roughness destroys the symmetry of the segments [38]. Therefore, experimental deviations from numerical solutions can be expected mainly due to fabrication defects (e.g., rough surfaces, rounded edges, curved surfaces, rough gaps, asymmetric dimers, heterogeneous nanowires shape, etc.) associated with the e-beam lithography and evaporation techniques resolution limits, which keep us away from achieving the ideal geometry proposed by simulations.

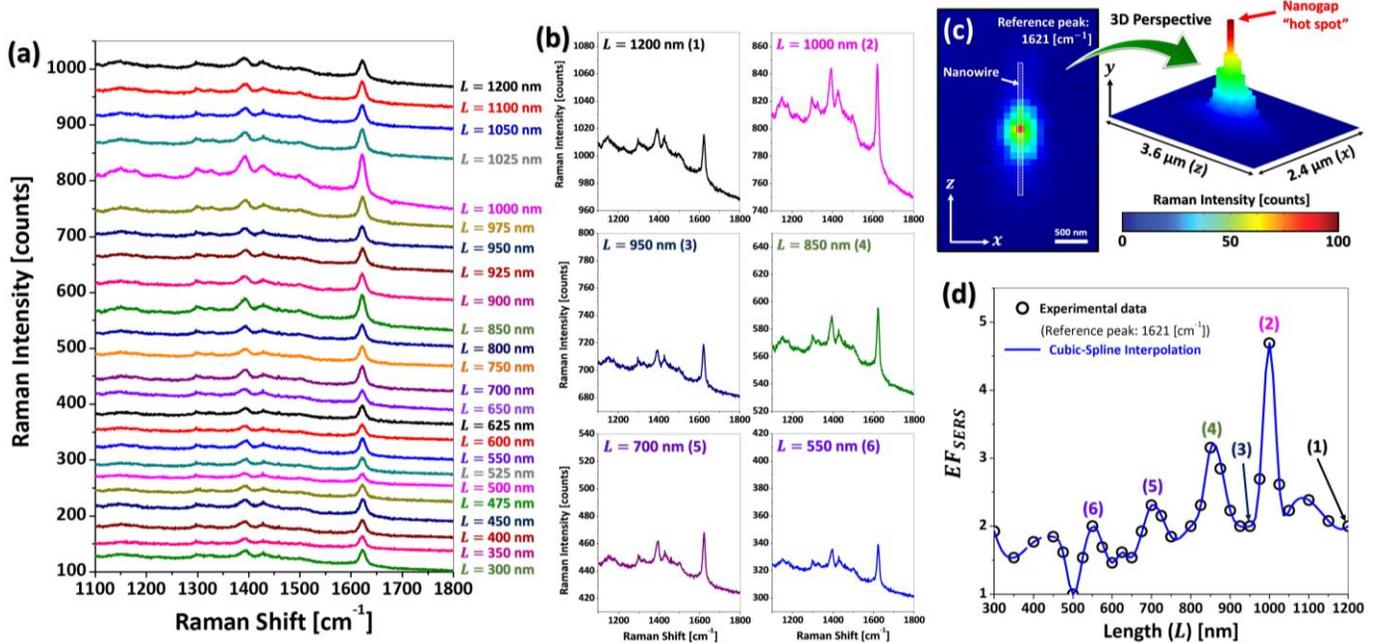

**Fig. 4. (a)** Averaged Raman spectra measured using the punctual acquisition mode, for $\theta = 0$ [deg]-laser polarization (633 [nm] wavelength), in MB-functionalized nano-gapped Au nanowires with lengths $L$ from 300 up to 1200 [nm]. Each spectrum corresponds to an average of 15 measurements carried out at the gap region in 15 different randomly chosen nanowires, with a total acquisition time of 30 seconds, according to the experimental conditions stated in section 2.2. **(b)** Magnification of some Raman spectra shown in (a) for better visualization of the main MB peaks manifested with different $L$ segments. **(c)** Submicron Raman map for $\theta = 0$ [deg] of a nano-gapped Au nanowire with $L = 1000$ [nm] using a power of 0.4 [mW] (10% of the maximum output power), 2 accumulations and 20 [s] of acquisition time per each pixel (size = 100 [nm]). The intensity scale was set taking as reference the most intense peak typically observed at 1621 [cm$^{-1}$]. **(d)** Magnitude of $EF_{SERS}(L)$ calculated for any $L$ value considering the criterion described by **Eq. (3)**. Additional $L$ conditions to those shown in (a) were analyzed to better define the behavior of the $EF_{SERS}(L)$ function. Numbered ($EF_{SERS}$; $L$) points and colors correspond to the same $L$, numbers and colors of spectra shown in **(b)**.

### 3.3. The key role of the laser polarization orientation on the SERS-based molecular sensing efficiency.
It is well known that the Raman signal enhancement is strongly dependent on the excitation wavelength, which is mainly determined by Electromagnetic Mechanisms (EMs) associated with the plasmonic-active nanostructure optical properties (excitation of LSPR states) and Chemical Mechanisms (CMs) associated with charge transfer interactions between the target molecule and this structure [72 – 74]. On the other hand, the laser polarization orientation can be a determining factor on the electric-near field enhancement, especially in plasmonic nanoantennas with high degrees of shape anisotropy like nanowires that can lead to the emergence of strong preferential SPPs states along different anisotropy axis [27]. Analogously to the excitation wavelength dependency, the effects of laser polarization can be also mediated by EMs and CMs. Chemical ones are mainly associated with the intrinsic Raman scattering cross-section dependency on light polarization due to the natural preferential spatial orientation of chemical bonds [1 – 3]. For its part, Electromagnetic ones are mainly associated with polarization conditions that maximize the efficiency of the light coupling with strong SPPs states that can be excited only in preferential spatial orientations, becoming much more important than CMs in nanostructures with considerable shape anisotropy [46, 47]. Thus, the laser polarization orientation could be a relevant factor impacting the Raman scattering enhancement and hence the efficiency of our nanostructures for SERS-based molecular sensing.

To study the effects of the laser polarization orientation angle $\theta$ (as schemed in **Fig. 2(b)**), samples with nominal $L = 1000$ [nm] were chosen for presenting the best Raman signal enhancement. This criterion was arbitrarily chosen to minimize the acquisition times and maximize the Raman intensity and spectral resolution during this study. In a first approach, the effects of varying $\theta$ were assessed by carrying out submicron Raman mapping studies with pixel size small enough (100 [nm]) to achieve a reasonably good resolution of the expected Raman intensity "hot spot" at the nanogap neighborhood. **Fig. 5(a)** shows the results obtained from this study for a MB-functionalized nano-gapped Au nanowire with nominal $L = 1000$ [nm] for different angles $\theta$ of 0, 30 and 90 [degree]. Here, each map was measured using the same experimental conditions stated for the case of the map shown in **Fig. 4(d)**, excepting the angle $\theta$. In a quick view, we can realize the important influence of the laser polarization orientation on the Raman signal enhancement at the nanogap. The intensity of the Raman signal at this zone tends to decrease as the angle $\theta$ increases, i.e., as the laser polarization alignment rotates away from the nanowire long axis. This observation is consistent with other similar results reported by H. Wei *et al* **[46]** and Hsu *et al* **[47]** who studied the polarization dependence of SERS signal in coupled gold nanoparticle-nanowire systems and AlN nanowires, respectively. In both studies, it was observed that the Raman signal enhancement was maximized when the polarization is aligned with the nanowire long axis, being consistent with the maximization of the light and SPPs states coupling under this alignment condition.

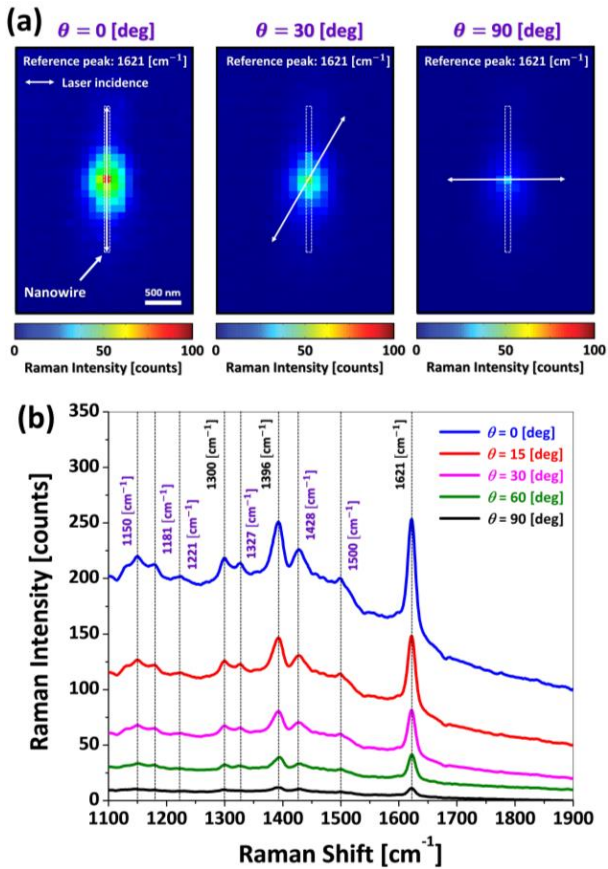

**Fig. 5. (a)** Submicron Raman maps (633[nm]-laser, pixel size = 100 [nm]) of a MB-functionalized nano-gapped Au nanowire with nominal $L = 1000$ [nm] for different laser polarization orientation angles ($\theta$) of 0, 30 and 90 [degree]. Each map was measured using the same experimental conditions stated for the case of the map shown in **Fig. 4(d)**, excepting the angle $\theta$. **(b)** Raman spectra measured for different $\theta$ values and $L = 1000$ [nm] by using the punctual acquisition mode. Each spectrum corresponds to an average of different spectra measured at a fixed $\theta$ in 15 different nanowires with nominal $L = 1000$ [nm]. Different Raman shift positions are highlighted to show the main vibrational modes typically observed for MB.

To quantify the effects of varying the angle $\theta$ on the SERS-based molecular sensing efficiency of our samples, Raman studies were carried out by using the punctual acquisition mode as close as possible to the nanowires' nanogap neighborhood as exemplified in **Fig. 2(b)**. This procedure is similar to that used for the case of the measurements shown in **Fig. 4(a)**. **Fig. 5(b)** shows the Raman spectra measured for MB-functionalized samples with nominal $L = 1000$ [nm] and different $\theta$ of 0, 15, 30, 60 and 90 [deg]. Each spectrum corresponds to an average of different spectra measured at a fixed $\theta$ in 15 different nanowires. For each nanowire spectrum, 10 accumulations and 5 [s] of acquisition time per accumulation were set. In this figure, different Raman shift positions are highlighted to show the main vibrational modes typically observed for MB **[53, 61 – 63]**. Consistently with the results shown in section 3.2, for the case of $\theta = 0$ [deg], it is observed the presence of the three main peaks of MB at 1300, 1396 and 1621 [cm$^{-1}$]. Additionally, other MB characteristic peaks were also observed. Two main peaks were identified at 1150 and 1181 [cm$^{-1}$], which are commonly associated with in-plane bending of C-H bonds and stretching of C-N bonds or higher-energy modes associated with C-H in-plane bending **[53, 62, 63, 75]**. Other vibrational modes at 1327, 1428 and 1500 [cm$^{-1}$] are recognizable from the Raman spectrum, which can be associated with in-plane ring deformation of C-H (1327 [cm$^{-1}$]), asymmetrical stretching of C-N bonds (1428 [cm$^{-1}$]) and asymmetrical stretching of C-C bonds (1500 [cm$^{-1}$]) **[53, 62, 63, 75]**. Therefore, under optimum measurement conditions ($\theta = 0$ [deg]), the nano-gapped Au nanowires ($L = 1000$ [nm]) reported here enable to achieve a good enough spectral resolution (low noise/signal ratio) to easily identify the vibrational spectrum of the analyte and hence its characteristic chemical footprint.

A competitive sensitivity for SERS-based molecular sensing usually ranges in the order of $10^0 - 10^2$ [ng/cm$^2$] **[76 – 85]**. On the other hand, considerably lower sensitivities have been reported in a few situations, achieving magnitudes about $10^{-2} - 10^{-1}$ [ng/cm$^2$] **[78, 82]** or even lower, about $10^{-3}$ [ng/cm$^2$] **[84]**. In the current case, good spectral resolution can be achieved for a relatively low surface mass density of MB (about 13 [ng/cm$^2$]) and fast acquisition times (< 1 min **[76, 78, 79, 81, 83, 84, 86 – 88]**), which represents a competitive performance for molecular sensing. Now, if the detection sensitivity is considered only taking into account the sensing active area (a laser spot size = 4 [μm$^2$] in the most pessimistic case), we can realize that the nanoantennas are capable of considerably amplifying the Raman signal of analyte mass ($m$) as low as $m = (13$ [ng/cm$^2$]$)(4$ [μm$^2$]$) = 52 \times 10^{-8}$ [ng] $= 5.2 \times 10^{-17}$ [g].

While $\theta = 0$[deg] is the optimum measurement condition in this kind of system, we can realize that the effects of increasing this angle can become critical and negatively affect the Raman signal enhancement. **Fig. 5(b)** demonstrates this fact, showing the systematic decrease of the Raman signal enhancement as the angle $\theta$ decreases. This means that the active nanostructure capability to induce efficient Raman scattering is decreasing as the laser polarization alignment rotates away from the nanowire long axis, worsening the spectral resolution of the analyte characteristic Raman peaks as observed. FDTD simulations have strongly suggested that the decrease of the Raman enhancement could be associated with a decrease of the ENFE at the nanogap zone as $\theta$ increases. **Fig. 6(a)** shows the calculated distribution of the normalized electric field magnitude ($E/E_0$) at the nanogap center in the $x - y$ plane (parallel to the nanowire lateral cross-section) and at the nanogap zone in the $x-z$ plane (perpendicular to the nanowire lateral cross-section) in a nano-gapped Au nanowire with $g = 22$ [nm], $L = 1000$ [nm] and cross-section of 74 [nm] x 74 [nm] for different angles $\theta$. These dimensions were chosen considering the actual experimental parameters obtained from the fabrication procedure. Additionally, substrate influence (Si with a 4 [nm]-thick native oxide layer) was considered in this case to have a more realistic approach. Results show that the electric-near field amplification ($E/E_0$) at the nanogap zone systematically decays as the polarization orientation angle ($\theta$) increases, which could be associated with lower efficiency of the light and SPPs states coupling when polarization is not aligned with the nanowire long axis in which strong propagating SPPs can exist [46, 47].

To quantify the effects of $\theta$ on the ENFE magnitude, we have used the same criterion presented in section 3.1, i.e., the figure of merit $EF'$ (**Eq. 2**). Similar to that case, we have considered the nanowire cross-section area (74 [nm] x 74 [nm]), as highlighted in the $x-y$ plane maps shown in **Fig. 6(a)**, to carry out the $EF'$ calculation. **Fig. 6(b)** shows the evolution of $EF'$ concerning $\theta$ variations from 0 to 360 [deg]. Results show a periodic dependence of $EF'$ observing well-defined enhancement maximum and minimum for $\theta = n\pi$ ($n \in \mathbb{Z}_0^+$) and $\theta = n\pi/2$ ($n = 2k+1$; $k \in \mathbb{Z}_0^+$), respectively. A critical dependence of $EF'$ with the angle $\theta$ can be observed, being consistent with those conclusions obtained from experimental Raman studies.

The consistency between the $\theta$-dependence predicted by numerical simulations for the electric-field enhancement and that observed by experimental studies for the Raman signal enhancement would suggest that the electromagnetic enhancement mechanism modulated by laser polarization orientation angle is a first-order parameter determining the SERS performance. This correlation is more evident if we compare and analyze the relative effects of angle $\theta$ on the $EF'$ (electromagnetic enhancement mechanism) and $EF_{SERS}$ (effective Raman enhancement). In this sense, **Fig. 6(c)** shows the normalized $EF'$ and $EF_{SERS}$ parameters taking as reference the highest value identified in each case for $\theta = 0$ [deg]. For $EF'$, this value is 159 according to **Fig. 6(b)**. For $EF_{SERS}$, calculated from results obtained for the same sample shown in **Fig. 5**, this value corresponds to 90 [counts]. $EF_{SERS}(\theta)$ was calculated using the same criterion presented by **Eq. (3)** in section 3.2. From this figure, we can realize that the theoretical model based on electromagnetic enhancement mechanisms represents a good approach for understanding the observed experimental tendency about the Raman enhancement dependency on the laser polarization orientation angle $\theta$.

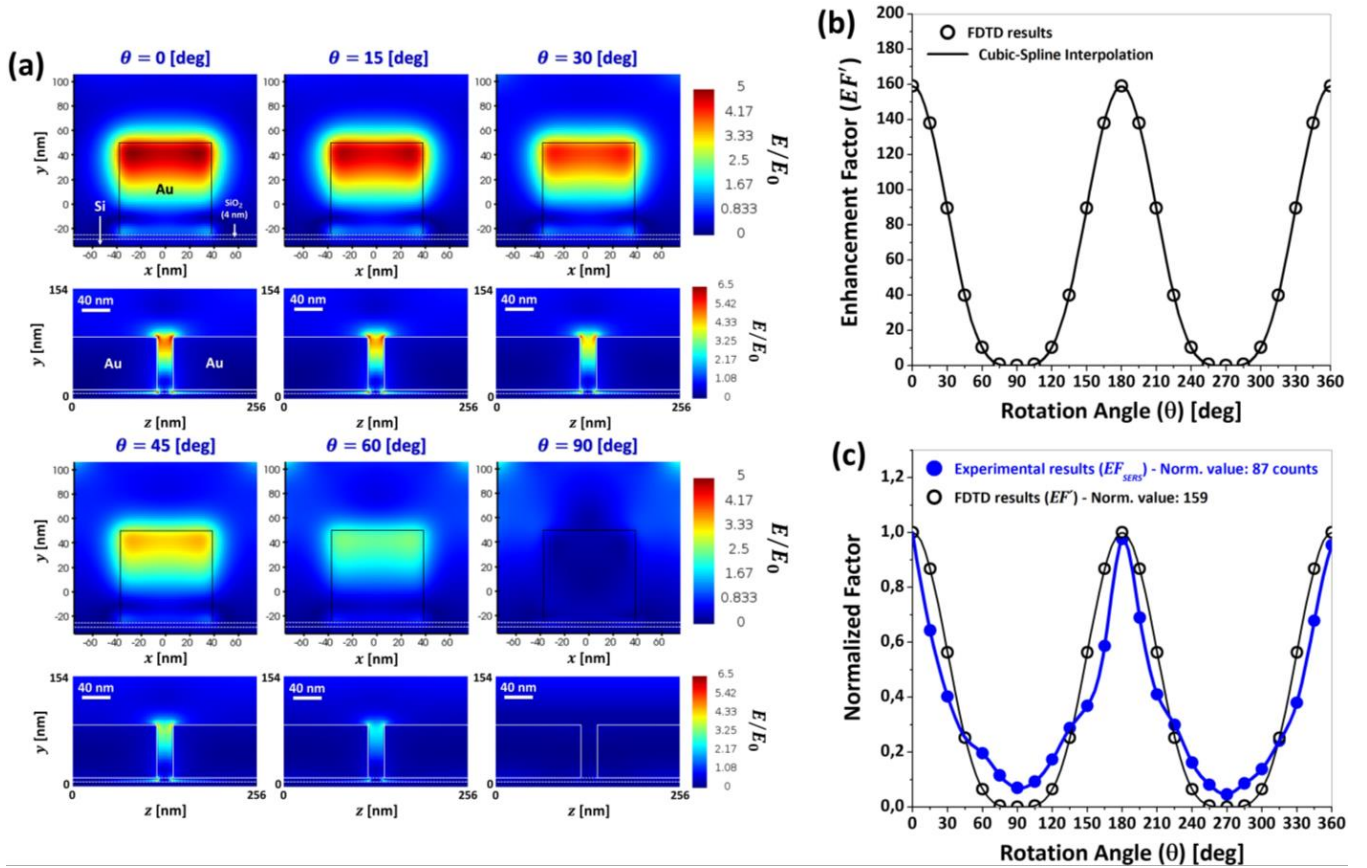

**Fig. 6. (a)** Distribution of the normalized electric field magnitude ($E/E_0$) at the nanogap center in the $x-y$ plane (parallel to the nanowire lateral cross-section) and at the nanogap neighborhood in the $z-y$ plane (perpendicular to the nanowire lateral cross-section) in a nano-gapped Au nanowire with $g = 22$ [nm], $L = 1000$ [nm] and cross-section of 74 [nm] x 74 [nm] for different 633 [nm]-laser polarization angles $\theta = 0, 15, 30, 45, 60$ and 90 [deg]. **(b)** Evolution of the $EF'$ concerning $\theta$ variations from 0 to 360 [deg] for the previous system using the criterion described by **Eq. (2)** in section 2.1. **(c)** Normalized $EF'$ and $EF_{SERS}$ taking as reference the highest value identified in each case for $\theta = 0$ [deg]. For $EF'$, this value is 159 according to **Fig. 6(b)**. For $EF_{SERS}$, calculated from results obtained for the same sample shown in **Fig. 5**, this value corresponds to 90 [counts].

While the values shown in **Fig. 6(c)** for theoretical and experimental results can differ, the curves' behavior is pretty similar. Of course, these deviations can be due to deviations from the ideal geometry proposed by the theoretical model. This fact is more evident if we analyze the behaviors around the minimum condition at $\theta = 90$ [deg]. While theory predicts $EF' = 0$ for $\theta = 90$ [deg] because of the non-coupling of light and SPPs states, the experimental results show that the Raman enhancement can be considerable under this angle condition, which demonstrates that other non-trivial electromagnetic enhancement mechanisms could be acting at the nanoscale. As suggested in section 3.2, an additional enhancement for $\theta = 90$ [deg] could be observed due to intrinsic nanometric roughness effects associated with the nanostructured granular nature of thin metallic films fabricated by e-beam evaporation (see **Appendix C**). Anyway, experimental results have demonstrated the critical role of the laser polarization orientation on the Raman signal enhancement in nano-gapped Au nanowires, also suggesting its potential key role in other kinds of plasmonic nanoantennas with high degrees of shape anisotropy. Thus, current results provide an interesting paradigm about the determining role of certain measurement conditions on the SERS-based molecular sensing efficiency in these kinds of nanostructures. In this case, the precise alignment of the laser polarization with the long axis of the sample is critical for maximizing the sensing efficiency.

**3.4. Performance of nano-gapped Au nanowires for SERS molecular detection.** The previously calculated $EF_{SERS}$ should be considered only as a demonstrative parameter, which is useful to study the relative effects of varying the nanowire length and laser polarization orientation on the SERS performance. To give a more accurate approximation to the "real" or absolute SERS enhancement factor of the studied nanostructure, it must be considered the weight of critical parameters like the number of SERS-active nanostructures per unit area ($\mu_M$), the surface density of the molecules adsorbed on the metal ($\mu_{mol}$), the metallic surface area of an individual nanostructure ($A_M$) and the effective area of the confocal scattering volume of the probing laser ($A_{eff}$). Considering these parameters, a well-accepted definition for the absolute enhancement factor ($EF_{SERS}^{abs}$) is described as follows [89]:

$$EF_{SERS}^{abs} = \frac{I_{SERS}}{I_{NR}}\left(\frac{N_{NR}}{N_{SERS}}\right) = \frac{I_{SERS}}{I_{NR}}\left(\frac{N_{NR}}{\mu_M \cdot \mu_{mol} \cdot A_M \cdot A_{eff}}\right), \qquad (4)$$

where $I_{SERS}$ and $I_{NR}$ are the signal intensities measured in SERS and Normal Raman (NR), respectively. $N_{SERS}$ and $N_{NR}$ represent the number of probed molecules contributing to SERS and normal Raman signals, respectively. These factors are modulated by the contribution of the $\mu_M$, $\mu_{mol}$, $A_M$ and $A_{eff}$ values. In general, all of these parameters are difficult to determine accurately, and defining them is not so direct in our case. Thus, we have proposed an alternative description for $N_{SERS}$ (which conserves the same physical meaning) based on different "area" parameters like $A_{EM}$ and $A_S$, which can be analytically estimate with relative ease. Making some reasonable assumptions, we can analytically estimate a more accurate value for $EF_{SERS}^{abs}$ considering the Surface Mass Density $\rho_{SM} = 13$ [ng/cm²] (or $13 \times 10^{-17}$ [g/µm²]) calculated in section 2.2 and the effective metallic area $A_{EM}$ of one nano-gapped Au nanowire covered by the active sensing area ($A_S = \pi r_{laser}^2 \cong 3.14$ [µm²]; $r$: laser spot radius) determined by the laser spot size (see **Fig. 2**). Considering $A_{EM}$ as the nano-gapped Au nanowire total surface area (area of the dimer covered by MB molecules) of the sample with the best performance ($L = 1000$ [nm]; $d = 74$ nm), we can estimate $A_{EM} = 2 \cdot$[one nanowire area] $= 2 \cdot [3(L \cdot d) + 2(d \cdot d)] = 2 \cdot [3 \cdot (1$ [µm])(0.074 [µm]) $+ 2 \cdot (0.074$ [µm])(0.074 [µm])] $= 0.465$ [µm²] (see scheme of **Fig. 3(a)** as reference). Thus, the number of probed molecules within the active sensing area contributing to SERS signal is determined as:

$$N_{SERS} = \frac{\rho_{SM} N_A A_{EM}}{m_M} \sim 1.2 \times 10^5, \qquad (5)$$

where $m_M = 320$ [g/mol] for MB and $N_A \cong 6 \times 10^{23}$ [molecules/mol] is the Avogadro number. To calculate $I_{NR}$ by confocal Raman spectroscopy measurements, 200 [µL] of a 1 [mM]-MB ethanolic solution was spread over a 25.4 [mm] x 12.7 [mm] non-plasmonically active soda-lime glass substrate (microscope slide). Then, the Raman spectra of this sample (normal Raman) were measured using the same experimental conditions indicated in section 2.2, aiming to establish a reliable comparison criterion between $I_{NR}$ and $I_{SERS}$. These intensities were calculated considering the most intense peak at 1621 [cm⁻¹] of the spectra shown in **Fig. 7(a)**, where the SERS spectrum corresponds to that shown in **Fig. 5(b)** for $\theta = 0$ [deg] and the normal spectrum corresponds to the above-mentioned sample. In both cases, the background was subtracted for a better visualization of the peaks absolute Raman intensity. Thus, $I_{NR} = 18$ [counts] and $I_{SERS} = 110$ [counts] were estimated from **Fig. 7(a)**.

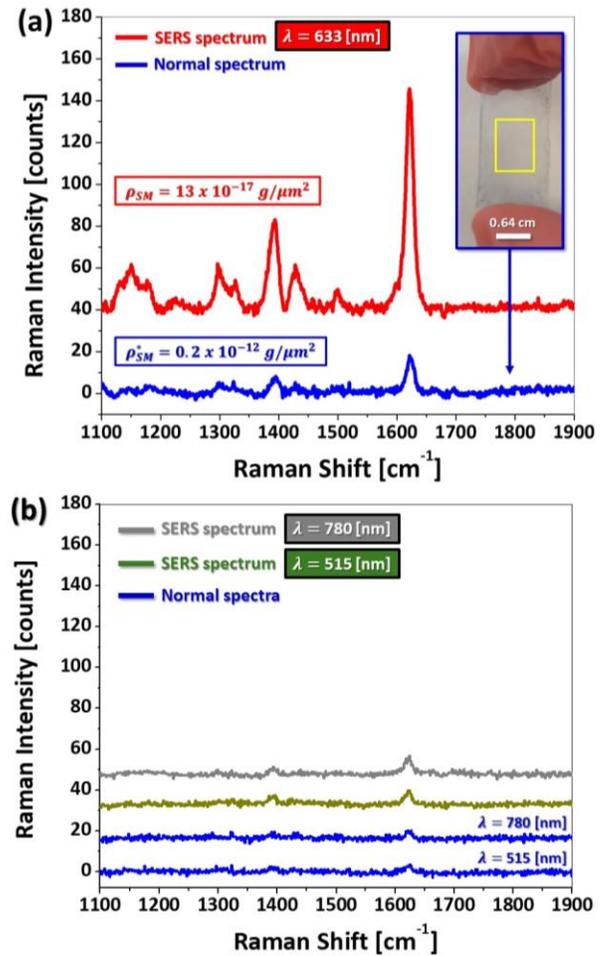

**Fig. 7.** (a) Raman spectra measured for nano-gapped Au nanowires (SERS spectrum, in red) and reference MB-covered soda-lime glass substrate (normal spectrum, in blue) for an excitation wavelength of $\lambda = 633$ [nm]. The SERS spectrum corresponds to that shown in **Fig. 5(b)** for $\theta = 0$ [deg], $d = 74$ nm, and $L = 1000$ [nm]. The normal spectrum corresponds to an average of 10 measurements carried out in different points on the sample surface. Each one of these measurements was carried out using the same experimental conditions indicated in section 2.2 for SERS-based ones. SERS and normal spectrum are associated with surface mass densities of $\rho_{SM} = 13 \times 10^{-17}$ [g/µm²] and $\rho_{SM}^* = 0.2 \times 10^{-12}$ [g/µm²], respectively. **Inset:** Image of the MB-covered soda-lime glass substrate used for normal Raman measurements. The yellow-delimited zone corresponds to the area considered for random punctual Raman measurements within that zone. **(b)** Same analysis carried out for the case of the figure (a) but considering other excitation wavelengths of 515 and 780 [nm]. The samples and all the other measurement conditions are the same as those stated for the case shown in (a).

Finally, assuming a homogenous distribution of the MB solution over the soda-lime glass substrate area (such that surface mass density for normal Raman $\rho_{SM}^* = 0.2 \times 10^{-12}$ [g/μm²]) and the effective sensed area as the active sensing area $A_S = 3.14$ [μm²], $N_{NR}$ is estimated using an analogous form of the **Eq. (5)** as follows:

$$N_{NR} = \frac{\rho_{SM}^* N_A A_S}{m_M} \sim 1.2 \times 10^9 . \quad (6)$$

Thus, the absolute enhancement factor achieved by the most efficient nano-gapped Au nanowires ($L = 1000$ [nm]) fabricated in this work is determined as follows:

$$EF_{SERS}^{abs} = \frac{I_{SERS}}{I_{NR}} \left( \frac{N_{NR}}{N_{SERS}} \right) = \frac{110}{18} \left( \frac{1.2 \times 10^9}{1.2 \times 10^5} \right) \cong 6.1 \times 10^4. \quad (7)$$

This experimentally determined SERS enhancement factor is significantly higher than those suggested by FDTD simulations (considering only electromagnetic mechanisms) for an optimum excitation wavelength of 633 [nm]. The additional enhancement could be due to chemical enhancement mechanisms associated with the high optical absorption of MB between 620 and 670 [nm] (with a peak at 665 [nm]) **[85]**, which would maximize the charge transfer efficiency between the plasmonic structure and excited molecule bonds for the used laser and hence the amplification of Raman intensity **[73, 74]**. This relation highlights the great relevance of tuning the laser source wavelength with the optical absorption band of the analyte, as well as with the optimum geometric parameter of the nanoantenna to obtain intense plasmonic resonance peaks of the metallic nanostructure, in order to achieve the maximum SERS detection efficiency. This fact can be observed by comparing **Figs. 7(a)** and **7(b)**, which demonstrates the considerably higher SERS enhancement achieved for the optimum excitation wavelength (633 [nm]) concerning other ones far from the analyte optical absorption range and nanowires plasmonic resonance, as suggested by preliminary simulations of the ENFE response of nano-gapped Au NWs with different lengths in the visible range (see **Fig. B** in **Appendix B**). In summary, the nanowires' geometrical configuration and laser wavelengths should be adequately chosen to optimize the SERS-based detection efficiency of a determined analyte considering its optical absorption properties.

The values typically reported for $EF_{SERS}^{abs}$ in single nano-gapped Au nanowires can vary in the order of $10^3 - 10^4$ **[37, 38, 90]** depending on the system's geometrical features. In this context, the enhancement factor achieved by the samples reported here shows a slight improvement of the SERS-based molecular sensing performance concerning similar reported systems. Other kinds of Au-based nanostructures fabricated by different techniques for SERS applications typically presents $EF_{SERS}^{abs} = 10^3 - 10^6$ **[91 – 97].** Thus, the nano-gapped Au nanowires studied in this work have demonstrated to be a promising plasmonic nanoantenna for fast and sensitive detection of relevant industrial-use molecules like MB, projecting future applications for competitive performance SERS-based molecular sensing.

## 4. Conclusion

Reproducible and reliable nano-gapped Au nanowires with nanogaps of $22 \pm 4$ [nm], $d = 74 \pm 3$ [nm], and lengths $L = 300 - 1200$ [nm] have been successfully fabricated by e-beam lithography and evaporation. The SERS-based molecular sensing efficiency of these nanostructures has been demonstrated to be considerably dependent on the excitation wavelength, $d$ and $L$ variations, as suggested by FDTD simulations that show the key role of these parameters in determining the electric-near field enhancement at the nanogap zone. Raman spectroscopy studies have demonstrated that the best Raman signal enhancement is achieved for nanoantennas with $d = 74 \pm 3$ [nm] and $L = 1000$ [nm] for a $g = 22 \pm 4$ [nm] and a 633 [nm] excitation laser, observing good spectral resolution for a relatively short acquisition time (30 [s]), while enabling to identify the main vibrational modes of the analyte and having a reliable representation of its characteristic chemical footprint. Polarization-dependent Raman studies in samples with $L = 1000$ [nm] have demonstrated the critical role of the laser polarization orientation angle $\theta$ on the Raman signal enhancement in nano-gapped Au nanowires, observing the systematic decay of this enhancement as $\theta$ increases and favors the misalignment of the polarization with the nanowire long axis. This fact could suggest the potential key role of this parameter in other kinds of plasmonic nanoantennas with high degrees of shape anisotropy. The consistency between the $\theta$-dependence predicted by numerical simulations for the electric-field enhancement and that observed by experimental studies for the Raman signal enhancement would suggest that the electromagnetic enhancement mechanism modulated by the polarization orientation is a first-order parameter determining the SERS performance. These results represent an interesting paradigm about the determining role of certain experimental conditions, like the nanoantenna precise geometry and the alignment of the laser polarization with the sample long axis, on the SERS-based molecular sensing efficiency in these kinds of nanostructures.

## Data availability statement

The datasets generated and analyzed during the current study are available from the corresponding authors on reasonable request.

## Declaration of competing interest

The authors declare that they have no known competing financial interests or personal relationships that could have influenced the work reported in this paper.

## Acknowledgements

S. R. thanks the *Consejo Nacional de Investigaciones Científicas y Técnicas* (CONICET, Argentina) for the financial support given through the program "Beca Interna Postdoctoral CONICET". T. K. acknowledges support from the European Union's Horizon 2020 research and innovation programme under the Marie Słodowska Curie Grant Agreement No. 101029928 (MANACOLIPO). M. L. P. thanks founding from CONICET (PIP 2022-2024), FONCyT (PICT-2020-SERIE A-02705) and UNCUYO (SIIP 2022 80020210100610UN) of Argentina. H. P. acknowledges support from the Science Committee of MESCS of Armenia (23RL-2A034).

## Appendix A: Raman spectra dispersion in samples with equal nominal $L$ value.

**Fig. A** shows an example of all the 15 Raman spectra measured for different nano-gapped Au nanowires samples with equal nominal $L$ values of 550 and 1000 [nm]. Each spectrum was measured following the experimental conditions previously stated in section 2.2. The spectra were artificially shifted in y-axis to better appreciate the experimental dispersion. In general, slight but considerable dispersion in the Raman intensity of spectra measured in samples with equal nominal $L$ values are observed. This could be associated with geometrical deviations from the nominal features due to the fabrication technique reproducibility and probable laser position shift effects during measurements that are critical at the nanoscale due to spatial resolution limits of confocal Raman microscopy. Nevertheless, this experimental dispersion is small enough to observe differences in the SERS performance attributed to variations in samples with different nominal $L$ values, as shown in the comparison of **Fig. A**.

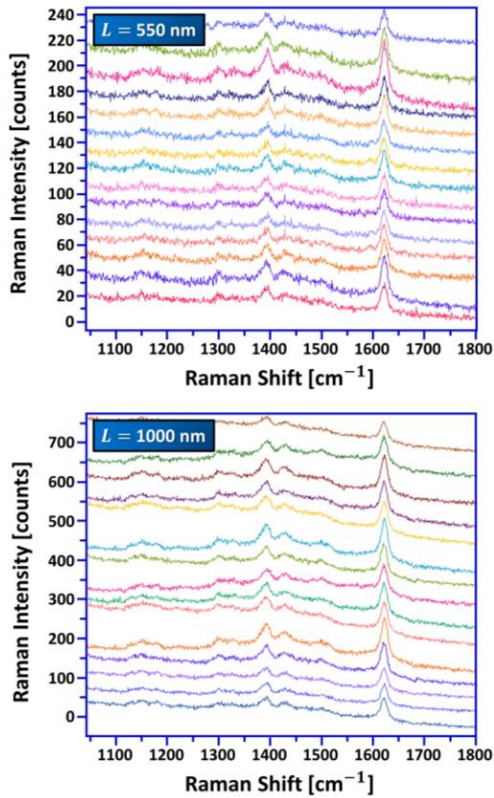

**Fig. A.** 15 Raman spectra measured for different nano-gapped Au nanowires samples with equal nominal $L$ values of 550 and 1000 [nm]. The spectra were artificially shifted in y-axis to better observe the experimental dispersion.

These results demonstrate that uncertainties in the laser positioning on nanogap are a secondary-order parameter affecting the Raman response deviations. This can be attributed to the larger laser spot size concerning position drift, nanowire width and nanogap size, which means that the laser dose on the nanogap neighborhood remains practically constant. Also, it evidences that the enhancement of the Raman scattering signal at the nanogap is predominant (note differences in relative intensity values) over any small deviation of the laser spot with respect to the gap position.

## Appendix B: Determining the optimum excitation wavelength for maximizing the NWs SERS performance.

**Fig. B** shows some representative results on the optical properties of the nano-gapped Au nanowires proposed in this study. **Fig. B(a)** shows the impact of different excitation wavelengths in the Visible-NIR range, typically used in Raman spectroscopy (455, 532, 633 and 780 nm), on the electric near-field enhancement ($E/E_0$) of a nano-gapped Au nanowire with $L = 1000$ [nm] ($g = 22$ [nm], $d = 74$ [nm]). In this case, the best enhancement is observed for an excitation wavelength of 633 [nm]. This enhancement condition was observed to be similar for any $L$ value. **Fig. B(b)** shows some examples of the $EF'$ obtained in the excitation range $400 - 800$ [nm] for different $L$ values, typically observing enhancements peaks in the range $625 - 640$ [nm] (close to 633 [nm]) with narrow bandwidths. In particular, these enhancement peaks can be associated with the excitation of the main plasmonic mode of nanowires (dipolar mode) **[27]**. A quick analysis of the reflectance spectra (**Fig. B(c)**) suggests that the geometry-induced enhancement can be associated with the emergence of higher-energy plasmonic resonance modes in the visible range (reflectance drops), which could be responsible for strong electric fields.

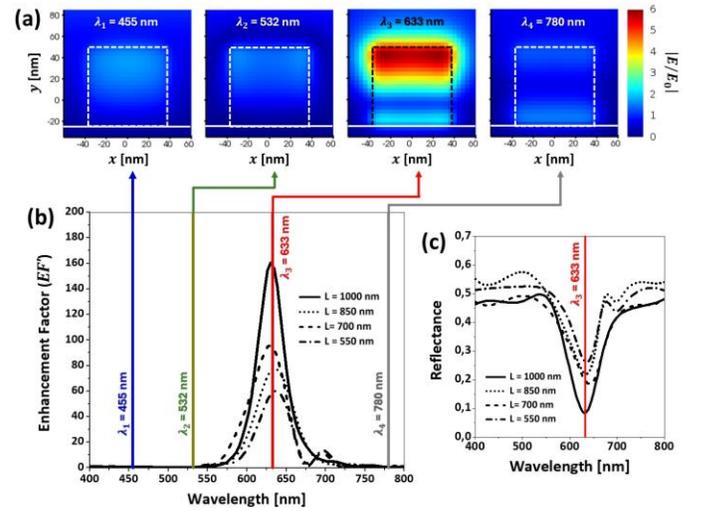

**Fig. B.** **(a)** Distribution of the normalized electric field magnitude ($E/E_0$) for $\theta = 0$ [deg] at the nanogap center in the $x - y$ plane (parallel to the nanowire lateral cross-section) in a nano-gapped Au nanowire with $g = 22$ [nm], $L = 1000$ [nm] and cross-section of 74 [nm] x 74 [nm] for different excitation wavelengths typically used in Raman spectroscopy **[8]**: $\lambda_1 = 455$ [nm], $\lambda_2 = 532$ [nm], $\lambda_3 = 633$ [nm] and $\lambda_4 = 780$ [nm]. **(b)** Evolution of the Enhancement Factor ($EF'$) determined by using **Eq. (2)** concerning different excitation wavelengths in Visible-NIR spectrum ($400 - 800$ [nm]) for nanowires with $L = 550, 700, 850$ and $1000$ [nm]. All nanowires have $g = 22$ [nm] and cross-sections of 74 [nm] x 74 [nm]. **(c)** Reflectance spectra associated with the cases previously mentioned.

**Appendix C: AFM studies of surface roughness and topography.**

AFM studies were carried out in both Au thin film deposited for future NWs fabrication and the final Au NWs sample. Films surface roughness ($R_a$) was observed to be about 1.7 [nm], and NWs surface roughness was observed to be similar reaching values in the order of 0.9 – 1.4 [nm]. Typical topologies observed in both cases are exemplified in **Fig. C** shown below, demonstrating the characteristic nanometric roughness.

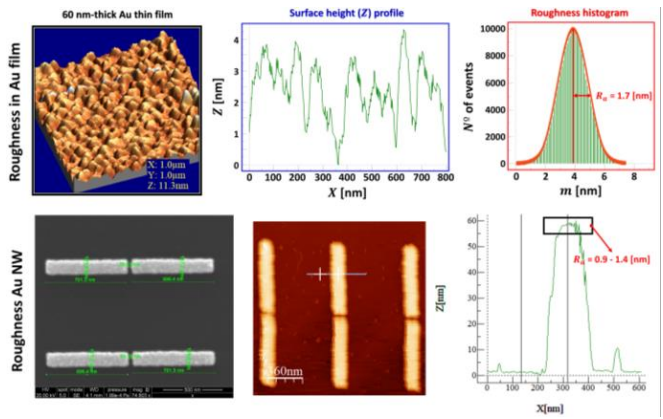

**Fig. C.** AFM studies of surface topography and roughness of an Au thin film deposited for future NWs fabrication and final Au NWs sample.